\documentclass[aip,jcp,reprint,superscriptaddress,citeautoscript,longbibliography,floatfix]{revtex4-1}

\usepackage{graphicx}
\usepackage{amsmath,amssymb}
\usepackage{bm}
\renewcommand{\epsilon}{\varepsilon}
\newcommand{\figurewidth}{0.9\columnwidth}
\newcommand{\narrowfigurewidth}{0.6\columnwidth}

\makeatletter
\renewcommand*{\NAT@spacechar}{~} 
\makeatother

\begin{document}

\title{Accurate and Efficient Numerical Simulation of Dielectrically
  Anisotropic Particles}

\author{Huanxin Wu}
\affiliation{Department of Physics and Astronomy,
    Northwestern University, Evanston, Illinois 60208, U.S.A.}

\author{Erik Luijten}
\email{luijten@northwestern.edu}
\affiliation{Department of Materials Science and Engineering,
    Northwestern University, Evanston, Illinois 60208, U.S.A}
\affiliation{Department of Engineering Sciences and Applied Mathematics,
    Northwestern University, Evanston, Illinois 60208, U.S.A}
\affiliation{Department of Physics and Astronomy,
    Northwestern University, Evanston, Illinois 60208, U.S.A.}

\begin{abstract}
  A variety of electrostatic phenomena, including the structure of electric
  double layers and the aggregation of charged colloids and proteins, are
  affected by nonuniform electric permittivity. These effects are frequently
  ignored in analytical and computational studies, and particularly difficult
  to handle in situations where multiple dielectric contrasts are present,
  such as in colloids that are heterogeneous in permittivity. We present an
  extension to the iterative dielectric solver developed by Barros and Luijten
  [\emph{Phys.\ Rev.\ Lett.\ \textbf{113}, 017801 (2014)}] that makes it
  possible to accurately compute the polarization of anisotropic particles
  with multiple dielectric contrasts. This efficient boundary-element
  method-based approach is applicable to geometries that are not amenable to
  other solvers, opening the possibility of studying collective phenomena
  of dielectrically anisotropic particles. We provide insight into the
  underlying physical reasons for this efficiency.
\end{abstract}


\maketitle

Electrostatic effects play a crucial role in colloidal suspensions, affecting
their stabilization, aggregation, and
electrokinetics.\cite{levin02,stradner04,linse99,morgan03} Computer
simulations can provide crucial insight in these electrostatic phenomena but
owing to computational limitations typically resort to coarse-grained
simulations, often using the so-called primitive model.\cite{luijten02a} This
model treats colloids and ions as discrete particles but the background
solvent as dielectric continuum. It is generally more accurate than mean-field
techniques, since fluctuations and steric effects (i.e., finite ion size) are
incorporated explicitly.  However, since a colloid typically has a different
electric permittivity than the surrounding solvent, it is also important to
account for induced surface (polarization) charge.  To resolve such effects in
the primitive model, the dielectric heterogeneity must be included when
solving Poisson's equation, which is typically analytically complicated and
numerically costly. Thus, polarization effects are ignored altogether in many
simulation models. Recent studies have demonstrated that this is not generally
justified, since dielectric effects can significantly alter the ionic density
profile near a surface\cite{messina02a,dossantos11,lue11,gan12,wu16,wu18b},
modulate ion mobility\cite{antila18a}, and affect the structure of
self-assembled aggregates\cite{barros14b}.  Until now, these studies have only
addressed dielectrically isotropic particles. In recent years, the study of
anisotropic particles has emerged as one of the frontiers in colloidal
science. These particles often display ``patchiness,'' i.e., surface regions
that possess distinct physical or chemical properties.  Such patchy particles
are promising candidates for drug delivery, molecular electronics,
self-healing materials, etc.~\cite{pawar10} However, accounting for dielectric
effects in these particles is considerably more complicated than for isotropic
spheres, since the standard image-charge techniques cannot be applied.

For dielectrically isotropic and homogeneous spherical colloids, the
traditional method of images is applicable to single-colloid systems, where
the image potential of an external charge is represented by the total
electrostatic potential of its Kelvin image and a line image
charge.\cite{neumann1883} For multiple colloids, generalizations of the
image-charge method through multi-level reflections~\cite{xu13a} and the
bispherical harmonic expansion method~\cite{rescic08} have been
proposed.\cite{gan16a} For colloids with \emph{anisotropic} dielectric
properties, the situation becomes exponentially more complicated, as there are
fewer symmetries to be exploited. More importantly, even in the simplest case
of piece-wise uniform dielectric domains, such anisotropic particles pose the
additional challenge of multiple dielectric mismatches.  Indeed, for more
complicated geometries, alternative approaches are preferred. Auxiliary-field
simulation methods, initially demonstrated as a Monte Carlo
algorithm\cite{maggs02} and subsequently adapted to molecular
dynamics,\cite{rottler04,pasichnyk04} offer the advantage that no explicit
solution of Poisson's equation is required. Owing to their local character,
they can be adapted immediately to systems with nonuniform
permittivity\cite{fahrenberger14a} and offer $\mathcal{O}(N)$ scaling.  Here,
we focus on the boundary-element method (BEM), in which sharp dielectric
interfaces are discretized into surface patches whose induced charge is found
through numerical solution of the integral form of Poisson's
equation.\cite{levitt78,hoshi87,bharadwaj95,allen01,
  boda04,tyagi10,jadhao12,lin14,barros14a} For complex geometries, the BEM
outperforms image-based approaches in efficiency and ease of
implementation.\cite{gan15} Its efficiency is dominated by the underlying
electrostatic solver and thus offers $\mathcal{O}(N \log N)$ or even
$\mathcal{O}(N)$ scaling. Auxiliary-field field methods are a natural choice
for problems with continuously varying
permittivities~\cite{fahrenberger14b,arnold15}, whereas systems with mobile
dielectric boundaries until now have only been demonstrated with a BEM-based
approach~\cite{barros14b}. For simple dielectric geometries (e.g., a single
isotropic sphere or planar interface), image-based methods offer superior
performance.

Although in principle BEM-based matrix equation solvers can be applied to
obtain the electrostatic potential around dielectric objects of arbitrary
geometry and dielectric configuration, their accuracy and convergence rate are
highly dependent on the conditioning of the boundary-element
equations.\cite{atkinson97} This conditioning depends not only on the BEM
formalism\cite{liang97}, but also on other factors, including object
geometry\cite{barros14a}, level of discretization, and shape of each
boundary-element.\cite{liang97,boschitsch02} \emph{Preconditioning} techniques
have been proposed in the context of both Poisson's equation\cite{buchau02}
and the Poisson--Boltzmann equation\cite{altman08} for multi-region dielectric
problems with large numbers of boundary elements, but neither the role of
dielectric heterogeneities (such as present in patchy colloids, proteins,
etc.)\ nor the spectrum of the BEM matrix have been examined explicitly.  We
perform such an analysis and attain an intuitive physical understanding of the
role of preconditioning, making it possible to extend the iterative dielectric
solver introduced in Refs.\citenum{barros14b} and\citenum{barros14a}---which
throughout this paper we will refer to as IDS---to achieve high accuracy and
fast convergence for systems of dielectrically heterogeneous particles.

Unlike finite-difference methods (FDM)\cite{gilson88a,gilson88b,baker01,yu07}
or finite-element methods (FEM),\cite{you93,holst00} which partition the
entire spatial domain, BEMs formulate partial differential equations as
boundary integral equations and only seek the boundary values.  For Poisson's
equation in electrostatics, the boundary values can either be the surface
charge density or the surface potential and its derivatives.
Since the permittivity often varies rapidly at dielectric boundaries, one
typically imposes sharp dielectric interfaces that separate piecewise uniform
media,\cite{levitt78} so that Poisson's equation only needs to be solved on
two-dimensional rather than three-dimensional (3D) grids.

We consider a dielectrically inhomogeneous system in space $\mathbf{V}$
consisting of piecewise uniform dielectric domains separated by smooth
boundaries $\mathbf{S}$, i.e., at arbitrary interface location~$\mathbf{s}$
with outward unit normal $\hat{\mathbf{n}}(\mathbf{s})$ we have different
relative permittivities $\epsilon_{\mathrm{in}}(\mathbf{s})$
and~$\epsilon_{\mathrm{out}}(\mathbf{s})$ on the opposing sides.  We assume
free charge distributions $\sigma_{\mathrm{f}}(\mathbf{s})$ on the interfaces
and $\rho_{\mathrm{f}}(\mathbf{r})$ in the bulk, which give rise to the induced
surface charge density~$\sigma_{\mathrm{pol}}(\mathbf{s})$,
for which various boundary integral representations have been derived.
Following Refs.\citenum{hoshi87} and\citenum{barros14a} we choose
\begin{equation}
  \bar\epsilon(\mathbf{s})\left[\sigma_{\mathrm{f}}(\mathbf{s})
    +\sigma_{\mathrm{pol}}(\mathbf{s})\right]
  +\epsilon_0\Delta\epsilon(\mathbf{s})\hat{\mathbf{n}}(\mathbf{s})
  \cdot \mathbf{E}(\mathbf{s})
  =\sigma_{\mathrm{f}}(\mathbf{s}) \;,
\label{eq:integral}
\end{equation}
where
$\bar\epsilon(\mathbf{s})=\left[\epsilon_{\mathrm{in}}(\mathbf{s})+
  \epsilon_{\mathrm{out}}(\mathbf{s})\right]/2$,
$\Delta\epsilon(\mathbf{s})=
\epsilon_{\mathrm{out}}(\mathbf{s})-\epsilon_{\mathrm{in}}(\mathbf{s})$,
and $\epsilon_0$ the vacuum permittivity.  The electric
field~$\mathbf{E}(\mathbf{s})$ comprises contributions from all (free and
induced) surface and bulk charges,
\begin{eqnarray}
  \mathbf{E}(\mathbf{s}) &=&
  \lim_{\delta\to 0}\iint\displaylimits_{\mathbf{S},|\mathbf{s}-\mathbf{s}'|>\delta}
  \frac{\left[\sigma_{\mathrm{f}}(\mathbf{s}')
  + \sigma_{\mathrm{pol}}(\mathbf{s}')\right] (\mathbf{s}-\mathbf{s}')}%
  {4\pi\epsilon_0|\mathbf{s}-\mathbf{s}'|^3} d\mathbf{s}' \nonumber\\
  && + \iiint\displaylimits_{\mathbf{V}\setminus\mathbf{S}}
  \frac{\rho_{\mathrm{f}}(\mathbf{r}')(\mathbf{s}-\mathbf{r}')}
  {4\pi\epsilon_0\epsilon(\mathbf{r}')|\mathbf{s}-\mathbf{r}'|^3} 
  d\mathbf{r}' \;,
\label{eq:efield}
\end{eqnarray}
where, to avoid the divergence of the layer potential, the infinitesimal disk
$|\mathbf{s}-\mathbf{s}'| \leq \delta$ is excluded.  $\epsilon(\mathbf{r}')$
is the relative permittivity at the off-surface
location~$\mathbf{r}'$. Equation~\eqref{eq:integral} relates the induced
charge density at surface location $\mathbf{s}$ directly to all other charges,
and thus has to be solved self-consistently.  To this end, the BEM discretizes
the interfaces and represents the continuous surface charge density
$\sigma(\mathbf{s})$ with a set of basis functions $f_i(\mathbf{s})$ defined
at each of $N$ boundary patches,
\begin{equation}
  \sigma(\mathbf{s})= \sigma_{\mathrm{f}}(\mathbf{s}) +
  \sigma_{\mathrm{pol}}(\mathbf{s})
  = \sum_{i=1}^{N} \sigma_i f_i(\mathbf{s})\;,
\end{equation}
where $\sigma_i$ is the weight at the $i$th patch.\cite{zauhar90} For
simplicity, piecewise-constant basis functions are widely
adopted,\cite{liang97,bardhan09a}
\begin{equation}
  f_i(\mathbf{s})=
        \begin{cases}
          1 & \text{if}\ \mathbf{s} \in s_i \\
          0 & \text{if}\ \mathbf{s} \not\in s_i
        \end{cases}\;,
\end{equation}
where $s_i$ is the enclosure of patch~$i$.  Under this approximation,
$\sigma(\mathbf{s})$ is discretized onto the $N$ boundary patches, each
carrying a charge density~$\sigma_i$.  For a finite number of patches, this
approximate $\sigma(\mathbf{s})$ does not satisfy Eq.~\eqref{eq:integral}
exactly, but results in a residual.  To minimize this residual, the BEM forces
it to be orthogonal to a set of test functions.\cite{gaul13} If these test
functions coincide with our basis functions, this approach reduces to the
standard Galerkin method.\cite{atkinson97} If, in addition to the
discretization, we assume that the bulk free charge distribution consists of
point charges, Eq.~\eqref{eq:integral} can be written in matrix form
$\mathcal{A}\bm{\sigma} = \mathbf{b}$, with
\begin{equation}
  \mathcal{A}_{ij} = \iint_{s_i}\left\{
    \bar{\epsilon}(\mathbf{s})\delta_{ij} + 
    \iint_{s_j}\left[\frac{\Delta\epsilon(\mathbf{s})}{4\pi}
      \frac{{\hat{\mathbf{n}}}(\mathbf{s})\cdot(\mathbf{s}-\mathbf{s}')}
      {|\mathbf{s}-\mathbf{s}'|^3}
    \right] d\mathbf{s}'\right\}d\mathbf{s}
\label{eq:A_ij}
\end{equation}
and
\begin{eqnarray}
  b_i &=& -\iint_{s_i}\left[\frac{\Delta\epsilon(\mathbf{s})}{4\pi}\sum_k
    \frac{q_k}{\epsilon(\mathbf{r}_k)}
    \frac{{\hat{\mathbf{n}}}(\mathbf{s})\cdot(\mathbf{s}-\mathbf{r}_k)}
    {|\mathbf{s}-\mathbf{r}_k|^3}
  \right] d\mathbf{s} \nonumber\\
  && + \iint_{s_i} \sigma_{\mathrm{f}}({\mathbf{s}}) d\mathbf{s}\;.
\label{b_i}
\end{eqnarray}
The nested integral in Eq.~\eqref{eq:A_ij}, if evaluated via one-point
quadrature at patch centroids, can lead to two different formulations.  If
$\mathbf{s}$ is evaluated at $s_i$, we have the collocation
approach,\cite{boda06} with
\begin{equation}
  \mathcal{A}_{ij} = \iint_{s_i}\bar{\epsilon}(\mathbf{s})
  \delta_{ij} d\mathbf{s} + 
  a_i\frac{\Delta\epsilon(\mathbf{s}_i)}{4\pi}\iint_{s_j}
  \frac{{\hat{\mathbf{n}}}(\mathbf{s}_i)\cdot(\mathbf{s}_i-\mathbf{s}')}
  {|\mathbf{s}_i-\mathbf{s}'|^3} d\mathbf{s}'\ .
\label{eq:A_ij_collocation}
\end{equation}
If $\mathbf{s}'$ is evaluated at $s_j$, we arrive at the qualocation
approach,\cite{tausch01} which at similar computational effort gives much
better accuracy,\cite{bardhan09a,bardhan09} especially for flat
patches.\cite{berti12}

For large-scale simulations, the solver must be not only accurate, but also
highly efficient. The IDS\cite{barros14a} takes the qualocation approach,
\begin{eqnarray}
  \mathcal{A}_{ij} &=& a_i\bar{\epsilon}(\mathbf{s}_i)\delta_{ij} + 
  a_j\frac{\Delta\epsilon(\mathbf{s}_i)}{4\pi}\iint_{s_i}
  \frac{{\hat{\mathbf{n}}}(\mathbf{s})\cdot(\mathbf{s}-\mathbf{s}_j)}
  {|\mathbf{s}-\mathbf{s}_j|^3} d\mathbf{s}\ ,
\label{eq:A_ij_new}
\\
  b_i &=& -\frac{\Delta\epsilon(\mathbf{s}_i)}{4\pi}\iint_{s_i}\sum_k
    \frac{q_k}{\epsilon(\mathbf{r}_k)}
    \frac{{\hat{\mathbf{n}}}(\mathbf{s})\cdot(\mathbf{s}-\mathbf{r}_k)}
    {|\mathbf{s}-\mathbf{r}_k|^3}
    d\mathbf{s} \nonumber \\
  &&  + \iint_{s_i} \sigma_f({\mathbf{s}}) d\mathbf{s}\;,
\label{eq:b_i_new}
\end{eqnarray}
where Eq.~\eqref{eq:A_ij_new} can be precomputed for fixed dielectric
geometries, but becomes time-dependent for mobile dielectric objects.  Thus,
to reduce computational cost, for $i\neq j$ the integral is approximated by
one-point (centroid) quadrature and for $i=j$ a curvature correction is added
by assuming disk-shaped patches with mean curvature.\cite{allen01} By further
assuming that source charges cannot approach the dielectric interfaces very
closely, and approximating Eq.~\eqref{eq:b_i_new} via one-point quadrature as
well, we arrive at simplified expressions for which the collocation and
qualocation approaches coincide,
\begin{eqnarray}
  \mathcal{A}_{ij} &=& \bar{\epsilon}_i\delta_{ij} + 
  a_j\frac{\Delta\epsilon_i}{4\pi}
  \frac{{\hat{\mathbf{n}}_i}\cdot(\mathbf{s}_i-\mathbf{s}_j)}
  {|\mathbf{s}_i-\mathbf{s}_j|^3}\ ,
\label{eq:A_ij_final}
\\
  b_i &=& -\frac{\Delta\epsilon_i}{4\pi}\sum_k
  \frac{q_k}{\epsilon(\mathbf{r}_k)}
  \frac{{\hat{\mathbf{n}}_i}\cdot(\mathbf{s}_i-\mathbf{r}_k)}
  {|\mathbf{s}_i-\mathbf{r}_k|^3}
  + \sigma_f({\mathbf{s}_i})\;,
\label{eq:b_i_final}
\end{eqnarray}
with $\bar\epsilon_i \equiv \bar\epsilon(\mathbf{s}_i)$,
$\Delta\epsilon_i \equiv \Delta \epsilon(\mathbf{s}_i)$, and
$\hat{\mathbf{n}}_i \equiv \hat{\mathbf{n}}(\mathbf{s}_i)$. To retain the
dimensionality of Eq.~\eqref{eq:integral}, we have divided both sides of
$\sum_j \mathcal{A}_{ij}\sigma_j=b_{i}$ by the patch area~$a_i$.  Instead of
solving this matrix equation through inversion of Eq.~\eqref{eq:A_ij_final},
the IDS\cite{barros14a} applies the iterative Generalized Minimal Residual
method (GMRES).\cite{saad86} One starts with an initial approximate
solution~$\bm{\sigma}^{(0)}$ and its corresponding initial residual
$\mathbf{r}^{(0)} = \mathbf{b}-\mathcal{A}\bm{\sigma}^{(0)}$.  Then, in the
$m$th iteration, a basis of the Krylov space is generated,
\begin{equation}
  \mathbf{K}^{(m)} = \mathrm{span}\{\mathbf{r}^{(0)},\mathcal{A}\mathbf{r}^{(0)},
  \ldots,\mathcal{A}^{m-1}\mathbf{r}^{(0)}\}\;.
\label{eq:krylov_basis}
\end{equation}
Since these basis vectors may be linearly dependent, Arnoldi iteration is
used to find orthogonal basis vectors
$\{\mathbf{q}_1, \mathbf{q}_2,\ldots,\mathbf{q}_m\}$, and
the $m$th approximate solution $\bm{\sigma}^{(m)}$ is obtained via linear
combination.
The computational cost of GMRES is dominated by the generation of each basis
vector in Eq.~\eqref{eq:krylov_basis}, which normally scales as
$\mathcal{O}(N^2)$.  However, our particular operator $\mathcal{A}_{ij}$
involves pairwise Coulomb interactions, so that the calculation can be
accelerated by a fast Ewald solver, such as the particle--particle
particle--mesh (PPPM) method\cite{hockney-eastwood-first,pollock96} at a cost
$\mathcal{O}(N\log N)$ or a Fast Multipole Method\cite{greengard97} at cost
$\mathcal{O}(N)$, without explicit matrix construction.  Once the surface
induced charge density is obtained, the electrostatic energy and forces follow
naturally, and can be used for Monte Carlo and molecular dynamics (MD)
simulations.

\begin{figure}[t]
  \centering
  \includegraphics[width=\narrowfigurewidth]{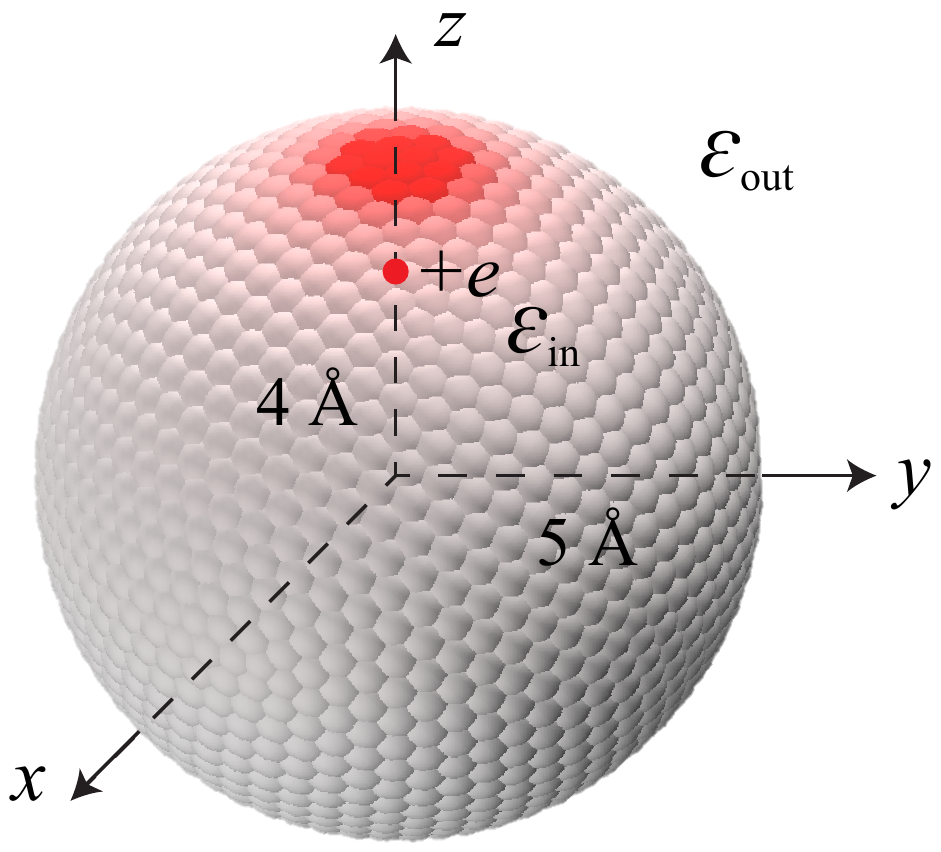}
  \caption{Test system introduced in Ref.\citenum{bardhan09} to examine the
    accuracy of various Poisson solvers.  A positive unit charge is placed
    4~\AA\ from the center of a dielectric sphere
    ($\epsilon_{\mathrm{in}}=80$) of radius 5~\AA\@. The sphere is embedded in
    a background medium with relative permittivity
    $\epsilon_{\mathrm{out}}=2$. Shading on the sphere surface indicates the
    induced charge.}
  \label{fig:berti_cmp_system}
\end{figure}

\begin{figure}[b]
  \centering
  \includegraphics[width=\figurewidth]{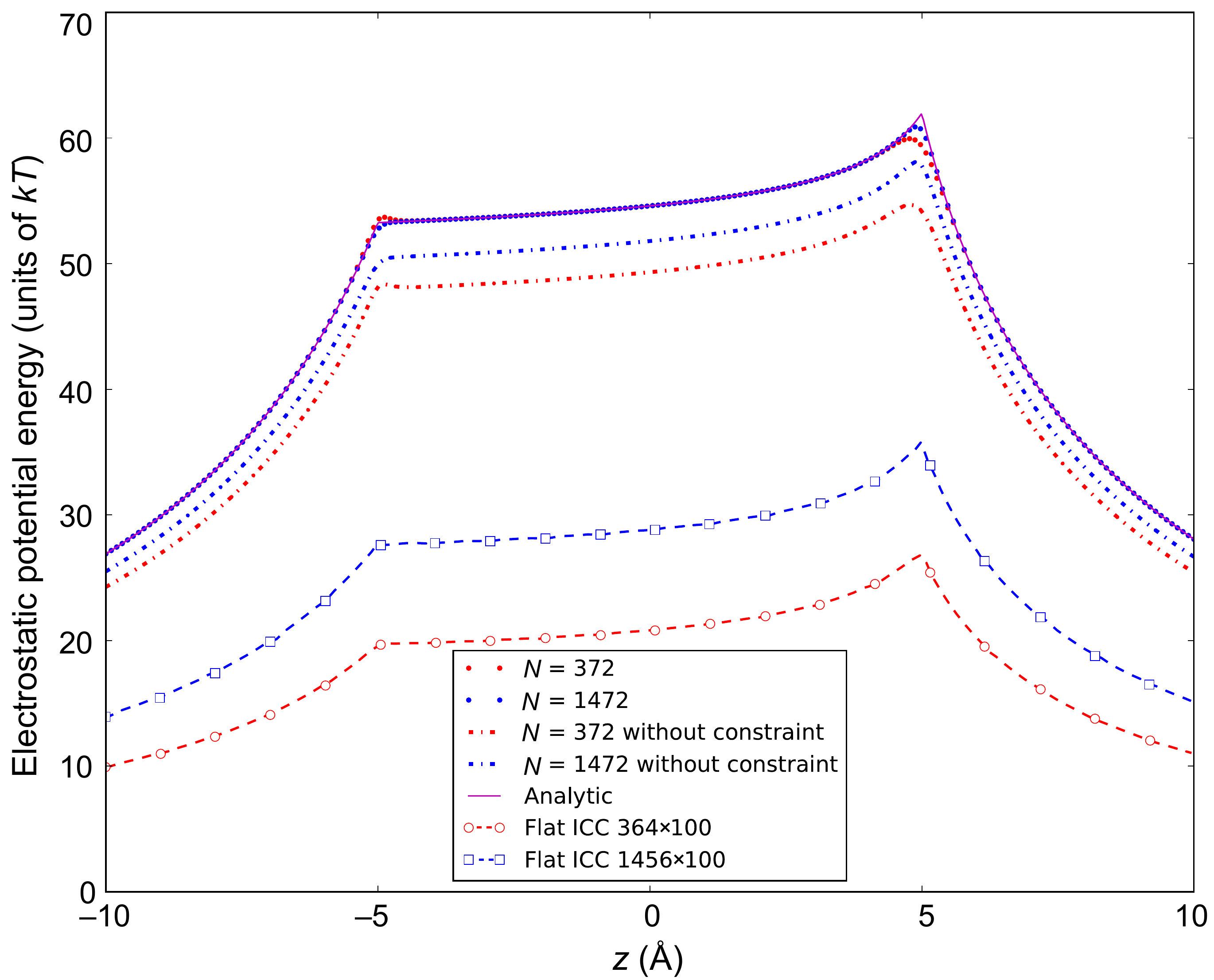}
  \caption{Accuracy comparison between the IDS (\emph{iterative dielectric
      solver}) implementation of Ref.\citenum{barros14a} and more costly
    alternative techniques for the induced charge potential along the $z$ axis
    for the configuration of Fig.~\ref{fig:berti_cmp_system}. Purple solid
    line marks the analytical solution. Open symbols represent data from
    Ref.\citenum{berti12} employing the induced charge computation (ICC) 
    method\cite{boda04}, where the
    sphere is discretized into $364$ (red open circles) and $1456$ (blue open
    squares) flat patches with $100$ subtiles per patch. The large discrepancy
    between these data and the analytical result can be significantly reduced
    by using one-point quadrature\cite{barros14a} (dashed lines marked
    ``without constraint,'' for comparable patch numbers, namely $372$ (red)
    and $1472$ (blue)).  Enforcing the net induced-charge constraint (cf.\
    Ref.\citenum{barros14a} and main text) improves the data (small red and
    blue solid circles) such that they become indistinguishable from the
    analytical result, except near the two surface boundary points
    ($z = \pm 5$\AA), where discretization effects dominate. This improvement,
    which is achieved at negligible additional computational cost, can be
    understood from the eigenvalue spectrum of the operator employed in the
    IDS (see main text).}
\label{fig:berti_cmp}
\end{figure}

In the context of the dielectrically heterogeneous particles examined below,
it proves insightful to first closely examine the performance of the IDS
approach proposed in Ref.\citenum{barros14a} for a uniform spherical particle,
with specific focus on the consequences of the one-point quadrature in Eqs.\
\eqref{eq:A_ij_final} and~\eqref{eq:b_i_final}. We adopt a test case from
Ref.\citenum{bardhan09}, i.e., the polarization potential of a dielectric
sphere ($\epsilon_{\mathrm{in}}=80$, $\epsilon_{\mathrm{out}}=2$) of radius
5~\AA, induced by a positive unit charge ($q=+e$) located \emph{inside} the
sphere, at a distance 4~\AA\ away from the sphere center
(Fig.~\ref{fig:berti_cmp_system}).  In Ref.\citenum{bardhan09}, this was found
to be a remarkably challenging system, with strong deviations between some
numerical approaches and the analytical solution~\cite{doerr04} for the
induced potential along the $z$-axis. The collocation approach~\cite{boda04}
was observed to yield a potential more than twice smaller than the analytical
result for a spherical surface discretized into 364 or 1456 flat tiles, with
each tile subdivided into $100$ elements for numerical integration
(Fig.~\ref{fig:berti_cmp}).  On the other hand, the qualocation
method~\cite{tausch01} was found to yield excellent agreement with the
analytical solution for the same tiling and subdivision.
Figure~\ref{fig:berti_cmp} shows that even the one-point quadrature
implementation of IDS performs far better than collocation with flat
disks\cite{boda04,berti12}, for similar global discretization levels (i.e.,
number of patches employed for the entire sphere).  Yet, the deviation from
the analytical result is still quite significant.  This is fully
mitigated by imposing the ``net induced-charge constraint'' derived in
Ref.\citenum{barros14a}.  For this test case, the net induced charge on the
sphere is nonzero,
and the total (free \emph{and} bound) charge should be
$q/\epsilon_{\mathrm{out}}$ (Ref.\citenum{barros14a}, Sec.~IV.H).  The total
bound charge itself consists of two contributions: the bound charge at the
source charge location ($q/\epsilon_{\mathrm{in}}-q$) and the surface induced
charge, so that the latter must equal
$q/\epsilon_{\mathrm{out}}-q/\epsilon_{\mathrm{in}}$.  To enforce this
physical constraint within GMRES, for simplicity we evenly distribute the net
charge over all patches for the initial trial solution~$\bm{\sigma}^{(0)}$ and
enforce the inner product of the patch areas $(a_1, a_2, \ldots, a_N)$ and
each subsequent basis vector,
$\Delta^{(m)} = \sum_{i=1}^N a_i(\mathbf{q}_m)_i$, to be zero, by subtracting
$\Delta^{(m)}/N$ from the computed induced surface charge of each patch at
every iteration. 
This technique, which comes at negligible computational cost, yields excellent
agreement with the analytical solution and rapid convergence as a function of
the number of surface patches.  Indeed, the accuracy is comparable with the
full qualocation approach at similar discretization levels, while avoiding the
use of subpatch discretization to obtain the second term of $\mathcal{A}_{ij}$
in Eq.~\eqref{eq:A_ij_new} (i.e., only a single evaluation per patch, rather
than numerical integration over $100$ subtiles). We note that
IDS\cite{barros14a} employs patches with a fixed curvature, implemented via a
curvature correction\cite{allen01}, but this is not to be confused with curved
surface elements,\cite{berti12} which are computationally far more
costly. Also, we have explicitly verified that this curvature correction has a
near-negligible effect on the results in Fig.~\ref{fig:berti_cmp}.

The high accuracy of IDS for this test case arises from two aspects of the
spectrum of the matrix operator $\mathcal{A}$. First, $\mathcal{A}$ is
well-conditioned.  For $\mathcal{A}\bm{\sigma}=\mathbf{b}$, the $L_2$-norm
condition number
$\kappa(\mathcal{A}) = \eta_\mathrm{max}(\mathcal{A}) /
\eta_\mathrm{min}(\mathcal{A})$
characterizes the sensitivity of the solution $\bm{\sigma}$ to a perturbation
in $\mathbf{b}$, where $\eta_\mathrm{max}(\mathcal{A})$ and
$\eta_\mathrm{min}(\mathcal{A})$ are the largest and smallest singular values
of $\mathcal{A}$, respectively. A perturbation $\delta\mathbf{b}$ in
$\mathbf{b}$ will lead to a perturbation $\delta\bm{\sigma}$ in $\bm{\sigma}$,
whose norm is bounded by the condition number,\cite{higham02}
\begin{equation}
  \frac{\left\lVert\delta\bm{\sigma}\right\rVert}%
  {\left\lVert\bm{\sigma}\right\rVert} \leq
  \kappa(\mathcal{A})
  \frac{\left\lVert\delta\mathbf{b}\right\rVert}%
       {\left\lVert\mathbf{b}\right\rVert} \;.
\end{equation}
A typical MD simulation employs a fast Ewald solver with moderate accuracy,
leading to inaccuracies in~$\mathbf{b}$.
Thus, an accurate solution of $\bm{\sigma}$ requires a small condition
number~$\kappa(\mathcal{A})$. For a normal matrix,
$\kappa(\mathcal{A}) = |\lambda_\mathrm{max}(\mathcal{A})|
/|\lambda_\mathrm{min}(\mathcal{A})|$,
with $\lambda$ its eigenvalues.
Whereas the sphere of Fig.~\ref{fig:berti_cmp_system} is dielectrically
isotropic, the patches differ slightly in area, causing the matrix
$\mathcal{A}$ to be asymmetric,
which results in complex eigenvalues, albeit with small imaginary parts.  The
condition number $\kappa(\mathcal{A})$ can be computed explicitly, since the
spectrum~$\lambda$ of~$\mathcal{A}$ was solved analytically for a spherical
geometry,\cite{barros14a}
\begin{equation}
  \lambda = \left\{\epsilon_\mathrm{out}, 
    \left(\frac{2}{3}\epsilon_\mathrm{out}+
    \frac{1}{6}\epsilon_\mathrm{in}\right),\ldots,
    \left(\frac{1}{2}\epsilon_\mathrm{out}+
    \frac{1}{2}\epsilon_\mathrm{in}\right)\right\}\;.
\label{eq:eigenvalues}
\end{equation}
yielding $\kappa(\mathcal{A}) = 41/2$, sufficiently small to guarantee a
well-conditioned matrix.

\begin{figure}
  \centering
  \includegraphics[width=\figurewidth]{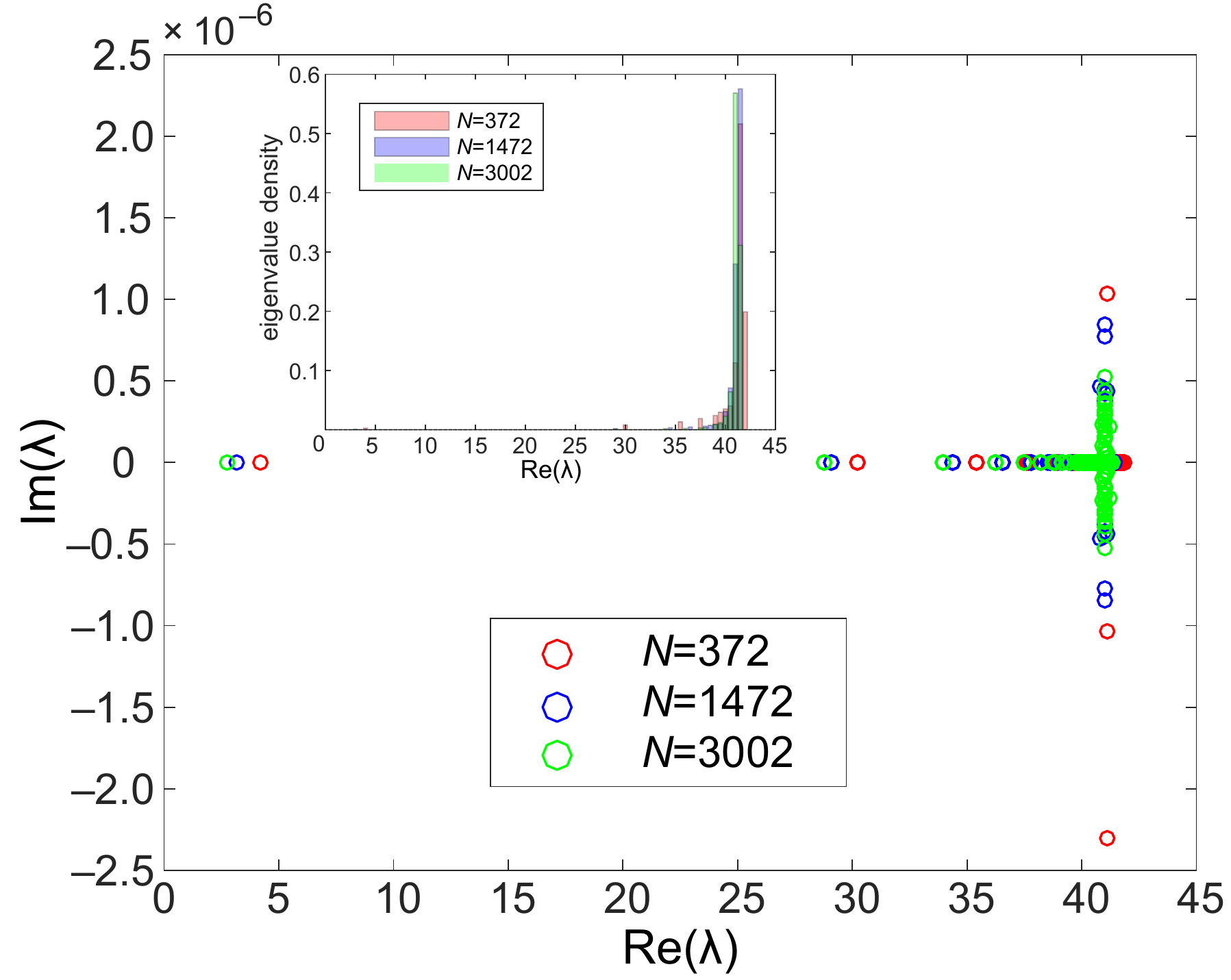}
  \caption{Complex spectrum~$\lambda$ of the operator $\mathcal{A}$ in
    Eq.~\eqref{eq:A_ij_final} for the dielectric sphere of
    Fig.~\ref{fig:berti_cmp_system} at different discretization levels~$N$. As
    $N$ increases, the real parts of the smallest and largest eigenvalues
    approach $2$ and~$41$, respectively, in agreement with
    Eq.~\eqref{eq:eigenvalues}\cite{barros14a}. The imaginary parts are small
    across the entire spectrum, reflecting the near-symmetric character
    of~$\mathcal{A}$. The inset shows the histogram of the real parts
    $\mathrm{Re}(\lambda)$, illustrating that apart from the outlying smallest
    eigenvalue all other values are clustered.}
  \label{fig:sphere_spectrum}    
\end{figure}

In Fig.~\ref{fig:sphere_spectrum} we evaluate the eigenvalues of $\mathcal{A}$
based upon Eq.~\eqref{eq:A_ij_final}, at different discretization levels.  The
extreme eigenvalues $\min(\lambda)$ and $\max(\lambda)$ gradually approach the
analytical predictions, i.e., $2$ and~$41$, as the patch number~$N$ is
increased.  The relative imaginary parts
${\mathrm{Im}(\lambda)}/{\mathrm{Re}(\lambda)}$ are indeed very small and
decrease as $N$ increases, indicating $\mathcal{A}$ is close to a normal
matrix. For $372$ patches, we find $\kappa(\mathcal{A})\approx 9.87$.

The second contribution to the accuracy of the IDS also follows from the
spectrum.  Namely, the convergence rate of GMRES depends on the eigenvalue
distribution of $\mathcal{A}$ in the complex plane.\cite{freund92} For fast
convergence, the eigenvalues should be clustered away from zero, i.e., the
distance between any two eigenvalues should be much smaller than the distance
of any eigenvalue from the origin.\cite{benzi02,larsen10}
Figure~\ref{fig:sphere_spectrum} shows that the minimum eigenvalue is isolated
from the other eigenvalues, compromising the quality of the spectrum.  The
eigenvector of this outlying eigenvalue is uniform, corresponding to a uniform
surface charge density.\cite{barros14a} As the total induced charge follows
from Gauss's theorem, this contribution can be computed analytically and
imposed as a constraint during the GMRES iterations.  Since $\mathcal{A}$ is
real and near-symmetric, its eigenvectors are orthogonal.  Thus, the physical
constraint imposed in the IDS precisely eliminates contributions of the
outlying eigenvalue. The remaining eigenvalues are clustered (cf.\
Fig.~\ref{fig:sphere_spectrum}, inset), ensuring fast convergence of the IDS
implementation in Fig.~\ref{fig:berti_cmp}.  Since each GMRES iteration
involves evaluation of the electric field at each patch location subject to
the accuracy of the Ewald solver, reduction of the number of iterations
reduces the cumulative error as well.

\begin{figure}
  \centering
  \includegraphics[width=\narrowfigurewidth]{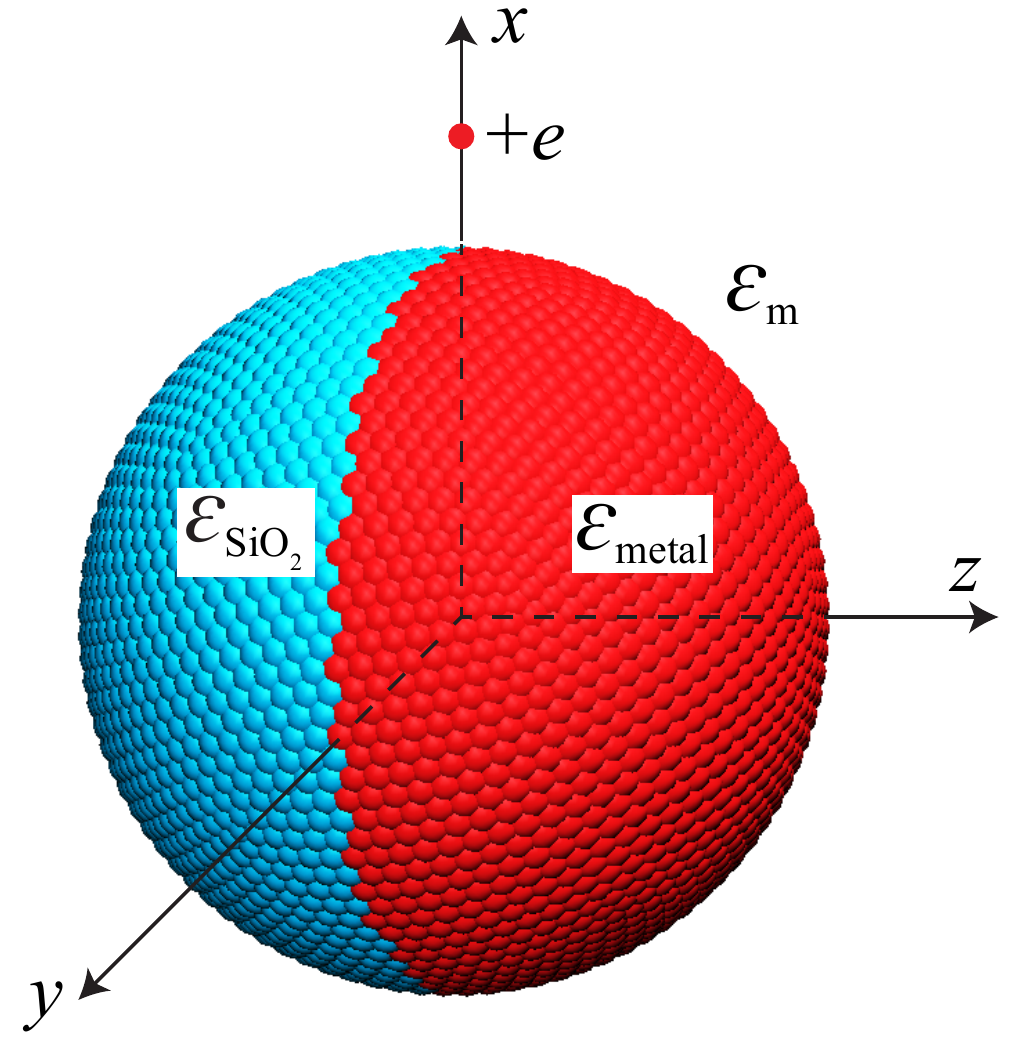}
  \caption{Silica/metal Janus particle of diameter~$14\sigma$ (with
    $\sigma=7.14$~\AA). The two hemispheres are separated by a disk-shaped
    dielectric interface (hidden from view). As a test case of the dielectric
    solver applied to dielectrically anisotropic particles, we examine the
    surface potential induced by a positive unit charge located at
    $(9\sigma,0,0)$.}
  \label{fig:janus_sketch}
\end{figure}

The IDS, including the net-charge constraint, has been successfully applied to
calculate the self-assembly and polarization of suspensions of binary mixtures
of isotropic spherical colloids.\cite{barros14b} Indeed, this solver is
applicable to arbitrary geometries, but the examination of the dielectric
sphere has shown that subtle issues may arise. To clarify these issues in the
case of dielectrically heterogeneous particles, where additional dielectric
interfaces arise, we consider the prototypical example of a Janus sphere
comprised of a silica hemisphere and a metallic hemisphere~\cite{wu16}. This
example exhibits three dielectric interfaces: two hemispherical surfaces and
one equatorial disk (Fig.~\ref{fig:janus_sketch}).  The silica side has
permittivity $\epsilon_{\mathrm{SiO_{2}}} = 4$ and the permittivity of the
conducting side is approximated by $\epsilon_{\mathrm{h}} = 10^5$.  The system
is embedded in an uniform dielectric medium representing water
($\epsilon_{\mathrm{m}} = 80$).  We set the diameter of the Janus particle to
$d = 14 \sigma = 10$~nm, where $\sigma=7.14$~\AA\ is the Bjerrum length.

To study the accuracy of the IDS, we compute the polarization charge induced
on a Janus sphere by a monovalent ion and compare the resulting surface
potential to a finite-element calculation performed using the COMSOL package
(Version~5.1, 2015).  The Janus particle has azimuthal symmetry about the
$z$-axis. The positive unit charge is placed~$9\sigma$ from its center, at a
polar angle $\theta=\pi/2$ (i.e., in the equatorial plane of the Janus
particle), so that the external source field acts equally on both hemispheres
(Fig.~\ref{fig:janus_sketch}).  Since the IDS yields the surface charge
density rather than the potential, additional errors are introduced when we
back-compute the potential on each surface patch, especially for the
contributions from immediately neighboring patches and from the patch itself.
To reduce such errors we adopt a mesh with 10\,242 patches on the sphere and
5\,000 patches on the equatorial disk. The electric field is evaluated via
PPPM Ewald summation, with a periodic simulation box that is large enough
($400\times400\times400\sigma^3$) to minimize periodicity artifacts. Both the
relative error of the Ewald summation and the convergence criterion of GMRES
are set to $10^{-6}$.  In the finite-element calculation, a ground potential
is imposed at the boundaries of the simulation box.  To suppress artifacts
resulting from this, we employ the same large simulation cell as for the
BEM-based calculation.  The entire 3D volume is discretized into a nonuniform
mesh with 3\,147\,897 tetrahedral elements.

\begin{figure}
  \centering
  \includegraphics[width=\figurewidth]{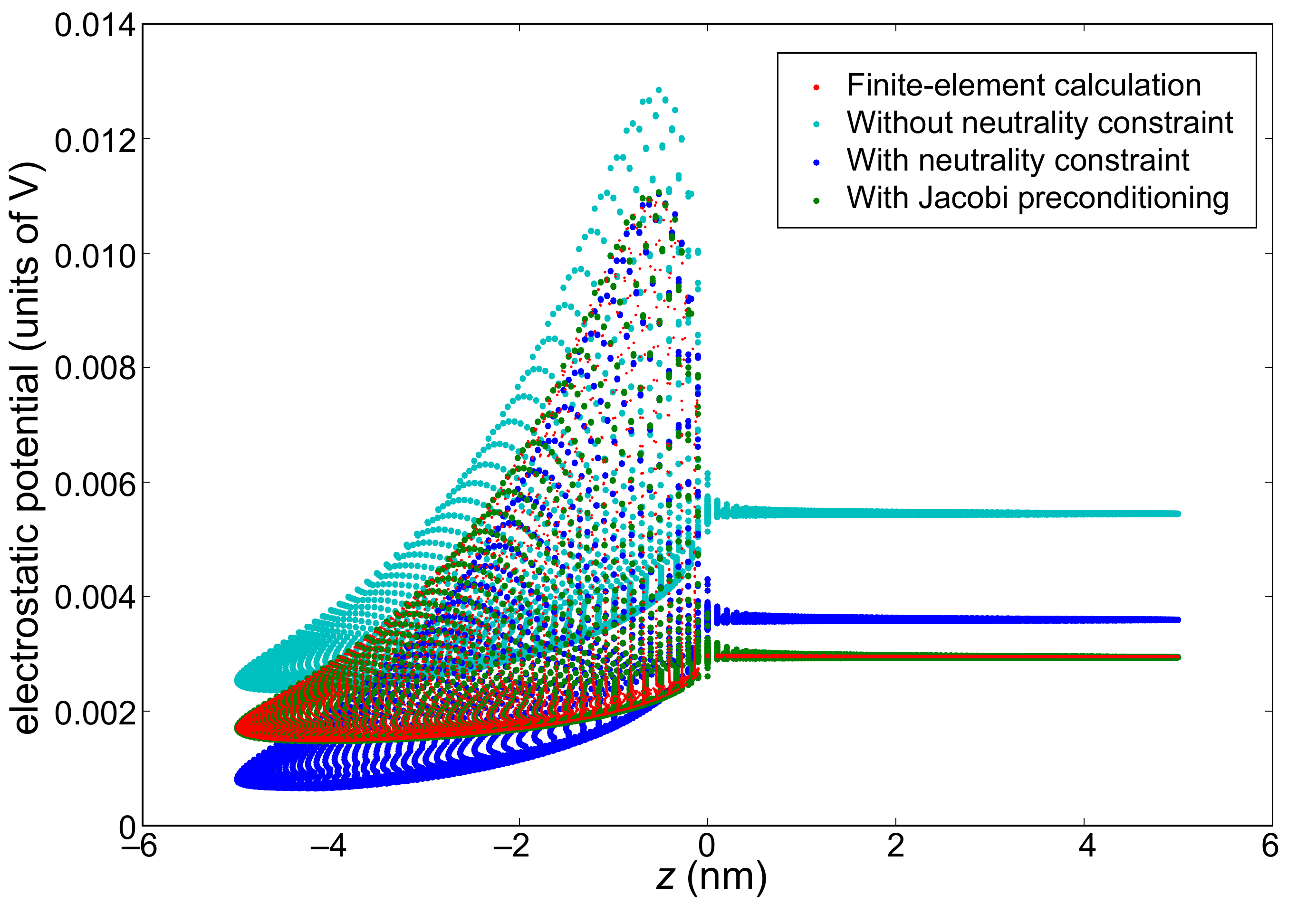}
  \caption{Comparison of different calculations of the total surface
    potential on a Janus sphere as set up in Fig.~\ref{fig:janus_sketch},
    i.e., a dielectric Janus particle embedded in water
    ($\epsilon_{\mathrm{m}}=80$), with a silica hemisphere
    ($\epsilon_{\mathrm{SiO_2}}=4$) and a metal hemisphere
    ($\epsilon_{\mathrm{h}}=10^5$), and a positive unit charge placed at
    $(9\sigma,0,0)$.  The spherical dielectric interface centered at the
    origin has radius $7\sigma = 4.998~\mathrm{nm}$ and is divided into
    10\,242 patches on the sphere and 5\,000 patches on the disk that
    constitutes the metal--silica interface. Red data points represent the
    surface potential as computed via a FEM calculation. The potential is
    constant on the metal hemisphere ($z>0$) and varies on the silica
    hemisphere ($z<0$). Cyan data are obtained with the IDS
    (iterative BEM-based dielectric solver) without any additional
    constraints. Blue data are obtained with the same solver, while
    constraining the net induced charge to zero. Both data sets exhibit
    significant deviations from the FEM solution. The green data points
    represent the IDS results obtained with a net-neutrality constraint as
    well as Jacobi preconditioning of the matrix operator. These results are
    obtained with negligible additional computational cost compared to a
    standard solver, and exhibit excellent agreement with the FEM data.  See
    main text for a detailed discussion.}
  \label{fig:comsol_cmp}
\end{figure}

Figure~\ref{fig:comsol_cmp} compares the two approaches for the total surface
potential.  We plot the potential at all patch centroids as a function of
their $z$ coordinates.  The red symbols show the FEM calculation, with a
constant potential on the metal hemisphere ($z>0$). The other symbols all
represent BEM calculations using the IDS,\cite{barros14a} with different
conditions. These data exhibit minor deviations from the constant potential
for small positive~$z$ (i.e., close to the equatorial plane), caused by
discretization. More important, however, are the systematic discrepancies.  If
no net induced-charge constraint is imposed, the BEM data (cyan) display a
strong, systematic deviation from the FEM data.  For the metal hemisphere, the
surface potential is almost twice higher than the correct result, and also for
the silica hemisphere the potential is consistently too high.  We emphasize
that the data have converged, but to the incorrect result.  This behavior is
similar to what we observed for the isotropic sphere
(Fig.~\ref{fig:berti_cmp}), although with significantly larger deviations.
Once the net induced-charge constraint is imposed (which amounts to a
net-neutrality constraint in this case, as the point charge is located outside
the sphere) for the entire Janus particle---i.e., for the entire system
comprised of the patches on the two hemispheres as well the patches at the
silica--metal interface---these deviations are significantly reduced, but by
no means negligible (Fig.~\ref{fig:comsol_cmp}, blue data). The potential on
the metal side is mostly constant, but still too high, and the potential on
the silica side only matches the FEM calculation close to the equator.

\begin{figure}
  \centering
  \includegraphics[width=\figurewidth]{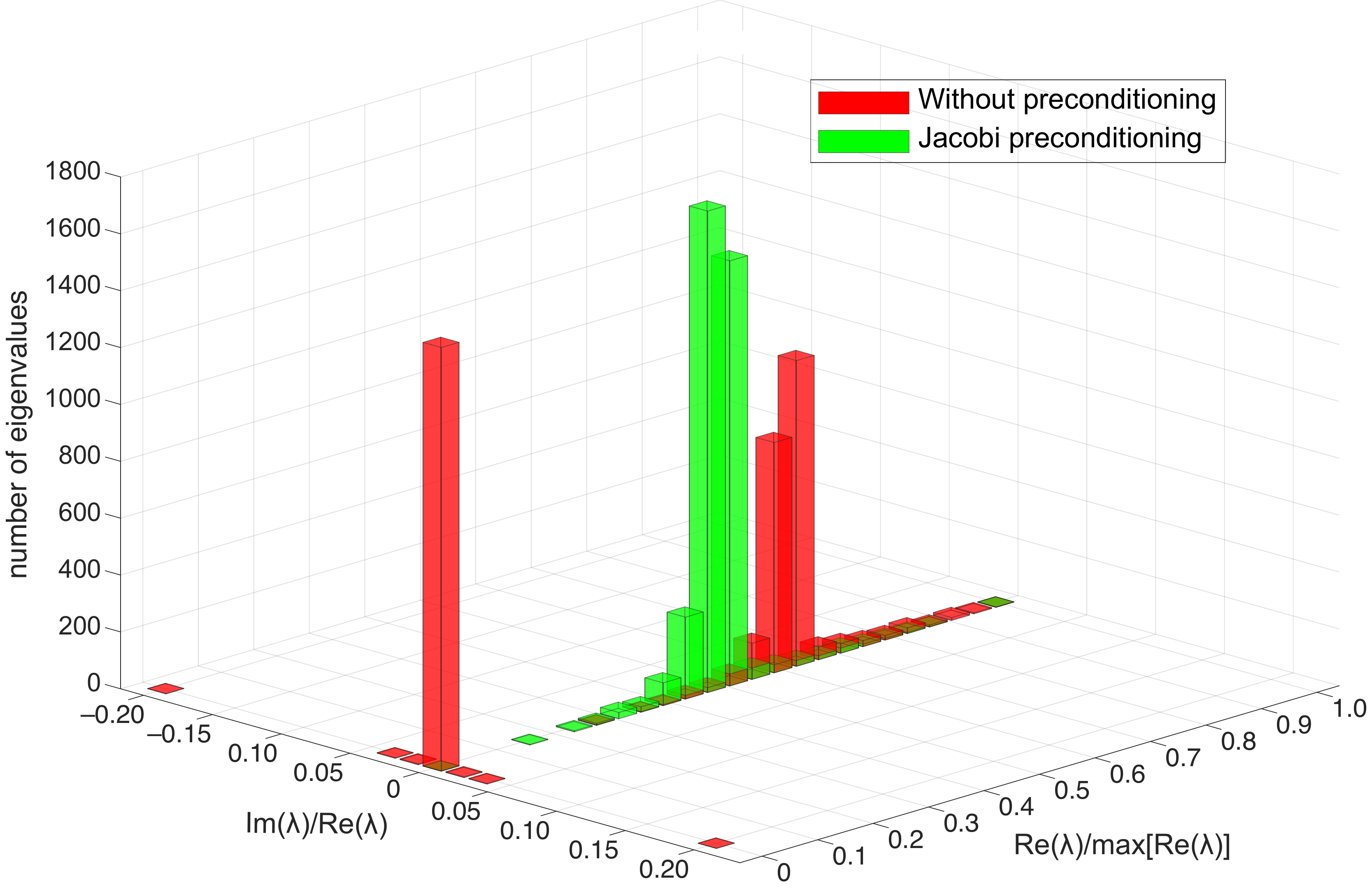}
  \caption{Comparison between the scaled spectra of the matrix $\mathcal{A}$
    for a Janus particle (Fig.~\ref{fig:janus_sketch}) with Jacobi
    preconditioning (green data) and without preconditioning (red data). The
    $x$-axis is scaled by the maximum of the real part of all eigenvalues. The
    $y$-axis gives the ratio between the imaginary and real parts of each
    eigenvalue.}
  \label{fig:janus_spectrum}
\end{figure}

To understand and resolve these discrepancies, we again turn to the spectrum
of the operator~$\mathcal{A}$. We find that the large dielectric mismatches at
the metal--water interface and the central metal--silica interface have a
detrimental effect on the condition number, yielding
$\kappa(\mathcal{A})=2.93\times10^3$.  Moreover, as illustrated in
Fig.~\ref{fig:janus_spectrum} (red data), the spectrum exhibits two groups of
normalized eigenvalues, clustered around 0 and 0.65, respectively, resulting
in slow convergence.  The anisotropy of the Janus particle also results in an
asymmetric matrix and significant imaginary parts for some of the eigenvalues,
hindering numerical solution of the matrix equation.\cite{saad89} To improve
this, we apply a preconditioner $\mathcal{M}$ to transform the matrix
equation,\cite{benzi02}
$\mathcal{M}^{-1}\mathcal{A}\bm{\sigma}=\mathcal{M}^{-1}\mathbf{b}$.  The
choice $\mathcal{M}=\mathcal{A}$ would yield perfect spectral properties, but
is prohibitively costly in situations where $\mathcal{A}$ is dynamic.
Instead, we observe that the simple Jacobi (or diagonal) preconditioner
$\mathcal{M} = \mathrm{diag}(\mathcal{A}) = \mathrm{diag}(\bar\epsilon_{ii})$
can be applied here. It is efficient for diagonally dominant
matrices,\cite{benzi02} as confirmed by the modified spectrum
(Fig.~\ref{fig:janus_spectrum}, green data).  With the Jacobi preconditioning,
the condition number drops 46-fold to 63.7, and the scaled eigenvalues are
clustered around 0.50. These improvements are reflected in the corresponding
results for surface potential (Fig.~\ref{fig:comsol_cmp}, green data), which
are in excellent agreement with the FEM calculations.  Intuitively, this
preconditioning remedies the disproportionate weight of patches with large
prefactors in Eq.~\eqref{eq:A_ij_final}, i.e., large $\bar\epsilon_i$ and
$\Delta\epsilon_i$ in the residual---precisely the situation that arises if
multiple dielectric mismatches are present.  This method of preconditioning
can be implemented in a particularly simple manner, namely in each iteration
of GMRES the residual of the $i$th patch is normalized by~$\bar\epsilon_{ii}$.

In summary, these results demonstrate that a combination of high accuracy in
the electrostatic summation, a strict convergence criterion in the GMRES
method, and a fine discretization level in the BEM are insufficient to
guarantee correctness of polarization charge calculations. However, with
proper preconditioning to reduce the matrix condition number for systems with
multiple dielectric contrasts and a physical (net induced-charge) constraint
to eliminate the effects of outlying eigenvalues in the operator spectrum, the
iterative dielectric solver of Ref.\citenum{barros14a} is capable of
accurately and efficiently resolving induced charge in systems with multiple
dielectric contrasts. A crucial observation is that the preconditioning
proposed here can be achieved at no additional computational cost. This is
essential for situations where the dielectric environment is time-dependent,
such as in dynamical simulations of colloids, proteins, etc., and thus the
induced charges must be resolved with the highest possible efficiency.  For
simplicity, we have focused on the prototypical Janus geometry. However, the
techniques presented here to improve the spectrum of the BEM matrix are
general and can be applied to a broad variety of dielectric
systems.\cite{han16a}

\begin{acknowledgments}
  The authors would like to thank Ming Han for FEM calculations and valuable 
  discussions. 
  This research was supported through award 70NANB14H012 from the
  U.S. Department of Commerce, National Institute of Standards and Technology,
  as part of the Center for Hierarchical Materials Design (CHiMaD), the
  National Science Foundation through Grant Nos.\ DMR-1121262 at the Materials
  Research Center of Northwestern University and DMR-1610796, and the Center
  for Computation and Theory of Soft Materials (CCTSM) at Northwestern
  University.  We thank the Quest high-performance computing facility at
  Northwestern University for computational resources.
\end{acknowledgments}

\end{document}